\documentclass[12pt]{iopart}

\usepackage{iopams, amssymb, amsfonts, graphicx}
\usepackage{color}

\renewcommand{\vec}[1]{\boldsymbol{#1}}
\newcommand{\matu}[1]{\boldsymbol{#1}}
\newcommand{\T}{\mathrm{T}}
\newcommand{\expt}[1]{\langle#1\rangle}
\newcommand{\im}{\mathrm{Im}}
\newcommand{\re}{\mathrm{Re}}

\begin{document}

\title[]{Heat transport in harmonic oscillator systems with correlated baths: Application to optomechanical arrays}

\author{Andr\'e Xuereb$^{1,2}$, Alberto Imparato$^3$, and Aur\'elien Dantan$^3$}

\address{$^1$\ Department of Physics, University of Malta, Msida MSD\,2080, Malta\\
$^2$\ Centre for Theoretical Atomic, Molecular and Optical Physics, School of Mathematics and Physics, Queen's University Belfast, Belfast BT7\,1NN, United Kingdom\\
$^3$\ Department of Physics and Astronomy, University of Aarhus, 8000 Aarhus C, Denmark}

\ead{andre.xuereb@um.edu.mt}

\begin{abstract}
We investigate the transport of phonons between $N$ harmonic oscillators in contact with independent thermal baths and coupled to a common oscillator, and derive an expression for the steady state heat flow between the oscillators in the weak coupling limit. We apply these results to an optomechanical array consisting of a pair of mechanical resonators coupled to a single quantised electromagnetic field mode by radiation pressure as well as to thermal baths with different temperatures. In the weak coupling limit this system is shown to be equivalent to two mutually-coupled harmonic oscillators in contact with an effective common thermal bath in addition to their independent baths. The steady state occupation numbers and heat flows are derived and discussed in various regimes of interest.
\end{abstract}

\vspace{2pc}

\maketitle

\section{Introduction}
Heat conduction in a physical system is a notoriously complex issue to investigate, as the dynamics depend strongly on the interaction between the system constituents as well as on the nature of the environmental baths and their coupling with the system~\cite{Dhar2008, Plenio2004, Galve2009, Galve2010, Asadian2013, Ghesquiere2013, Manzano2013, Mari2013}. However, low-dimensional systems in contact with different energy or particle baths represent an excellent test-bed for some of the most recent ideas in classical and quantum out-of-equilibrium statistical physics~\cite{Dhar2008}. For example, one can show that a chain of quantum harmonic oscillators in contact with two heat baths at different temperatures exhibits a steady state fluctuation theorem, setting constraints on the entropy production~\cite{Dhar2007}, in all respects equivalent to the fluctuation theorem for the corresponding classical case~\cite{Fogedby2014}. Furthermore, chains of oscillators have been used as model systems to study heat conduction in solids, in particular to test the validity of Fourier law, according to which the heat current across a material subject to a temperature gradient scales as the inverse of the system size~\cite{Dhar2008}. Motivated by the growing interest in the thermodynamic properties of out-of-equilibrium quantum systems~\cite{Kosloff2013, Strasberg2013, Esposito2010}, we investigate in this paper a prototypical system consisting of a set of quantum harmonic oscillators, each in contact with an independent thermal bath, and all coupled to a common oscillator, which is itself in contact with its own bath. The coupling to the common oscillator effectively mediates an interaction between the different oscillators and baths, which renders the description of the quantum dynamics quite complex in general. While one can envisage many situations that this system may model, our study is specifically motivated by opto- or electromechanical arrays~\cite{Eisert2004, Heinrich2011, Chang2011, Tomadin2012, Seok2012, Schmidt2012, Xuereb2012, Botter2013, Aspelmeyer2013}, in which electromagnetic radiation can affect the motion of mechanically compliant structures, thereby allowing effective transport of phonons between the mechanical elements~\cite{Xuereb2014}. In addition to their widespread use for sensing and for communication technologies, opto/electromechanical systems have made great progress towards operation in the quantum regime in the past decade~\cite{Aspelmeyer2013}. This has motivated, among other things, their potential application to quantum thermodynamics and the investigation of quantum heat engines, pistons, etc.~\cite{Mari2012,Xuereb2014,Zhang2014,Ian2014,Mari2014,Dechant2014}. Among their chief virtues is the highly tunable coupling with electromagnetic radiation which can be enhanced with a resonator and which allows for flexible engineering of interactions and readout. Arrays of mechanical oscillators are particularly interesting as long-range interactions between the mechanics can be engineered~\cite{Xuereb2012,Xuereb2013} in a well-controlled fashion and collective phenomena, such as self-oscillations, phonon lasing or synchronisation, can occur~\cite{Heinrich2011, Massel2012, Zhang2012, Ludwig2013, Bagheri2013, Kemiktarak2014, Schmidt2013, Chesi2014}.

\begin{figure}
\centering
\includegraphics{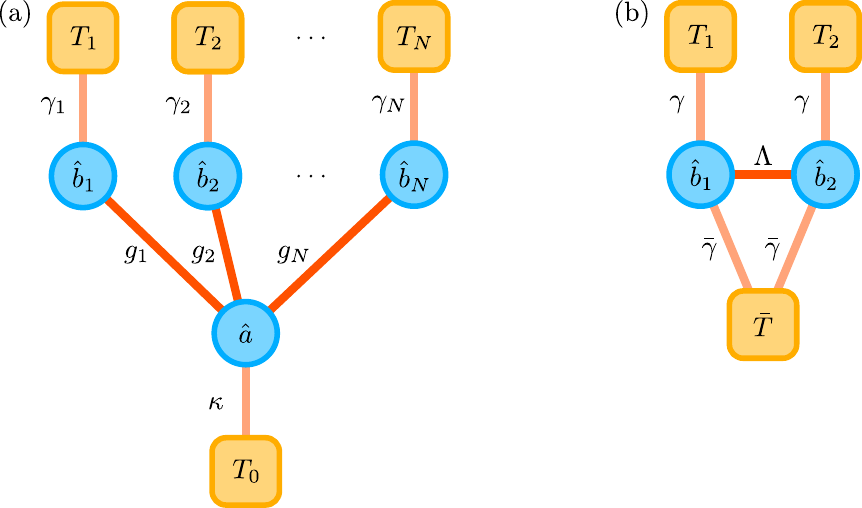}
\caption{(a)~$N$ oscillators, each in contact with its own independent thermal bath and coupled to a common oscillator, which is in contact with its own bath. (b)~Equivalent model system for an array with $N=2$: After adiabatic elimination of the common oscillator, the oscillators are effectively mutually coupled and in contact with both their initial thermal bath and a common bath.} \label{fig:Schematic}
\end{figure}

In this work we investigate phonon transport in an ensemble of identical oscillators in contact with independent thermal baths with (possibly) different temperatures and coupled to a common oscillator---a single electromagnetic field mode in an opto/electromechanical array setting---as illustrated schematically in Fig.~\ref{fig:Schematic}. We start by deriving a general expression, Eq.~\eref{eq:JlBare}, for the heat flow through the individual elements in steady state when the couplings to the common mode are weak and this mode can be adiabatically eliminated. After discussing and solving the special case where all the baths are at the same temperature (Sec.~\ref{sec:EqualTemp}), we consider the general problem consisting of a two-mechanical-resonator array coupled to one electromagnetic field and thermal baths with arbitrary temperatures. We show that, after the adiabatic elimination of the field, it is equivalent to a generic two-oscillator system with an effective mutual linear coupling, an effective common bath and two independent baths (Secs.~\ref{sec:OM} and~\ref{sec:OMIndependentCB}). We solve this generic problem for thermal Markovian baths and derive expressions, Eqs.~(\ref{eq:SteadyStateOcc}) and~(\ref{eq:SteadyStateHF}), for the steady state occupation and heat flows of the mechanics. In Sec.~\ref{sec:Discussion}, we discuss the results in various parameter regimes which could be realised through a suitable engineering of the optomechanical interaction and give an indication of the various systems that could be used for investigating the effects we explore. Our results show the possibility for engineering heat flow and the Fourier law in arrays of harmonic oscillators possessing only indirect coupling, and we illustrate this by proposing a practical application in optomechanical arrays. 
Systems with several oscillators,  where the interaction between the different constituents can be easily tuned, and the local temperature can be precisely controlled are highly desirable given the current interest in out-of equilibrium physics, and so the set-up we propose can represent an advancement with respect to the experimental set-ups which have been recently used to study the entropy production in systems in contact with only two heat baths \cite{Ciliberto13,Pekola13}. Likewise, our system can represent a test-bed to extend some of the concepts of stochastic thermodynamics \cite{Seifert} to the quantum case, allowing one to measure, e.g.,  the heat current, and thus the entropy production in a quantum out-of-equilibrium system.   Finally, we conclude in Sec.~\ref{sec:Conclusion} by surveying our results and putting them in the context of possible future work.

\section{Heat flow for $N$ oscillators coupled to a common oscillator and independent thermal baths}
We consider a system composed of $N$ identical harmonic oscillators, mutually uncoupled, but all linearly coupled to a common harmonic oscillator, and in contact with independent thermal baths. We denote by $\varrho$ the density matrix for the $(N+1)$-partite system. The master equation for $\varrho$ can be written ($\hbar=1$):
\begin{equation}
\dot\varrho=-i[\hat{H},\varrho]+\sum_{j=1}^N\mathcal{L}_j\varrho+\mathcal{L}_\mathrm{a}\varrho,
\end{equation}
where $\hat{H}$ is the Hamiltonian governing the evolution,
\begin{equation}\label{eq:Lj}
\mathcal{L}_j\varrho=\gamma_j(n_j+1)D[\hat{b}_j]\varrho+\gamma_jn_jD[\hat{b}_j^{\dagger}]\varrho,
\end{equation}
with
\begin{equation}
D[\hat{b}]\varrho=2\hat{b}\varrho\hat{b}^\dagger-\hat{b}^\dagger\hat{b}\varrho-\varrho\hat{b}^\dagger\hat{b}
\end{equation}
and $\hat{b}_j$ the annihilation operator of the $j$\textsuperscript{th} oscillator ($1\leq j\leq N$), $\gamma_j$ the coupling rate of the $j$\textsuperscript{th} oscillator to its bath, whose mean occupation number is $n_j$, and
\begin{equation}\label{eq:La}
\mathcal{L}_\mathrm{a}\varrho=\kappa D[\hat{a}]\varrho,
\end{equation}
with $\hat{a}$ the annihilation operator of the final harmonic oscillator, and $\kappa$ its coupling constant to its own bath, assumed at zero-temperature. We assume that $\hat{H}$ has the following form~\cite{Xuereb2014}:
\begin{equation}
\label{eq:StartingHam}
\hat{H}=\Omega\hat{a}^{\dagger}\hat{a}+\sum_{j=1}^N\omega\hat{b}_j^\dagger\hat{b}_j+\sum_{j=1}^Ng_j\bigl(\hat{a}+\hat{a}^\dagger\bigr)\bigl(\hat{b}_j+\hat{b}_j^\dagger\bigr),
\end{equation}
where $\omega$ and $\Omega$ are the oscillators' frequencies and $g_j$, assumed real, represents the coupling strength of the $j$\textsuperscript{th} oscillator to the final one. If $g_j$ is small compared to the other frequencies of the problem, $\hat{a}$ can be adiabatically eliminated from the dynamics.\footnote{The assumptions of the adiabatic elimination procedure are mainly (i)~that one subsystem evolves on a much faster time-scale than the rest of the system, (ii)~that the coupling between the two is weak, and (iii)~that the total density matrix is approximately a tensor product between the fast and slow subsystems.} For the interested reader, we reproduce this elimination procedure in Appendix I.

Let us indicate the reduced, $N$-partite, density matrix that results from this elimination process with $\rho$. The heat flow into or out of the $l$\textsuperscript{th} element\footnote{We use the convention where positive heat flow corresponds to heat flowing \emph{into} the system.} is given by~\cite{Asadian2013}
\begin{equation}
J_l=\Tr\bigl(\hat{H}\mathcal{L}_l\varrho\bigr).
\end{equation}
We now make use of our adiabatic elimination procedure to write, in steady state and to lowest order in the coupling constants $g_j$, $\varrho=\rho_\mathrm{ss}\otimes\rho_\mathrm{a}$, where $\rho_\mathrm{a}$ is the steady-state density matrix for the $(N+1)$\textsuperscript{st} harmonic oscillator. Let us first take the trace with respect to this mode\footnote{We use the notation ``$\neg\mathrm{a}$'' to refer to the set of all the modes other than mode `a.'}:
\begin{eqnarray}
J_l&=\Tr_{\neg\mathrm{a}}\Biggl(\Tr_\mathrm{a}\Biggl\{\Biggl[\Omega\hat{a}^{\dagger}\hat{a}+\sum_{j=1}^N\omega\hat{b}_j^\dagger\hat{b}_j+\sum_{j=1}^Ng_j\bigl(\hat{a}+\hat{a}^\dagger\bigr)\bigl(\hat{b}_j+\hat{b}_j^\dagger\bigr)\Biggr]\nonumber\\
&\hspace{10em}\times\mathcal{L}_l\rho_\mathrm{ss}\otimes\rho_\mathrm{a}\Biggr\}\Biggr)\nonumber\\
&=\Tr_{\neg\mathrm{a}}\Biggl\{\Biggl[\Omega\langle\hat{a}^{\dagger}\hat{a}\rangle+\sum_{j=1}^N\omega\hat{b}_j^\dagger\hat{b}_j+\sum_{j=1}^Ng_j\langle\hat{a}+\hat{a}^\dagger\rangle\bigl(\hat{b}_j+\hat{b}_j^\dagger\bigr)\Biggr]\mathcal{L}_l\rho_\mathrm{ss}\Biggr\}.
\end{eqnarray}
Following the usual methodology, the state described by $\rho_\mathrm{a}$ has been shifted to describe a zero-mean state, so that $\langle\hat{a}+\hat{a}^{\dagger}\rangle=0$; furthermore, the contribution due to $\langle\hat{a}^{\dagger}\hat{a}\rangle$ disappears due to the trace-preserving nature of the superoperator $\mathcal{L}_l$. The adiabatic elimination used to derive the above expression assumes a weak coupling between the two subsystems. Consistently with this approximation, we keep our results at lowest order in the coupling strength and neglect the effects of correlations between mode `a' and the other modes. We then obtain
\begin{equation}
J_l=\Tr\Biggl\{\omega\sum_{j=1}^N\hat{b}_j^\dagger\hat{b}_j\mathcal{L}_l\rho_\mathrm{ss}\Biggr\},
\end{equation}
where we drop the subscript from the trace because there is no longer any ambiguity. By exploiting the bosonic commutation relations and the cyclic property of the trace, we obtain an explicit expression for the steady-state heat flow:
\begin{equation}
J_l=2\omega\gamma_l[n_l-\Tr(\hat{b}_l^\dagger\hat{b}_l\rho_\mathrm{ss})]=2\omega\gamma_l\bigl(n_l-\langle\hat{b}_l^\dagger\hat{b}_l\rangle\bigr).
\label{eq:JlBare}
\end{equation}
The work in this section extends the treatments of Refs.~\cite{Asadian2013} and~\cite{Nicacio2014} to the case where the $N$ oscillators do not interact directly, but via coupling to a common, adiabatically eliminated, mode. To proceed further, one needs an expression for the average occupation of the mechanical elements. As we shall now illustrate, there are various situations under which we can calculate this quantity explicitly.

\section{Case of identical baths}
\label{sec:EqualTemp}
It is possible to obtain an analytic expression for the heat flow in the case in which the different thermal baths to which the elements are connected are identical, i.e., they are characterised by one single coupling constant $\gamma_j=\gamma$ and occupation number $n_j=n$. To continue, it is convenient to introduce a normal-mode basis. Define $g:=\sqrt{\sum_{j=1}^Ng_j^2}\neq 0$, $\vec{g}^{(1)}=(g_1/g\ g_2/g\ \dots\ g_N/g)^\T$, and $\vec{g}^{(l)}$ ($2\leq l\leq N$) such that the set $\{\vec{g}^{(l)}\}$ forms an orthonormal basis.\footnote{Since we define $g_j$ such that they are all real, we can assume that the entire basis set is composed of entirely real vectors.} We use this ``collective'' basis to define a new set of $N$ normalised harmonic oscillator annihilation operators $\tilde{b}_j$ ($1\leq j\leq N$):
\begin{equation}
\left(\begin{array}{c}
\tilde{b}_1\\
\tilde{b}_2\\
\vdots\\
\tilde{b}_N
\end{array}\right)=\left[\begin{array}{c}
{\vec{g}^{(1)}}^\T\\
{\vec{g}^{(2)}}^\T\\
\vdots\\
{\vec{g}^{(N)}}^\T
\end{array}\right]\cdot\left(\begin{array}{c}
\hat{b}_1\\
\hat{b}_2\\
\vdots\\
\hat{b}_N
\end{array}\right)=\matu{G}\cdot\left(\begin{array}{c}
\hat{b}_1\\
\hat{b}_2\\
\vdots\\
\hat{b}_N
\end{array}\right),
\end{equation}
whereby
\begin{equation}
\dot\varrho=-i[\hat{H},\varrho]+\sum_{j=1}^N\tilde{\mathcal{L}}_j\varrho+\mathcal{L}_\mathrm{a}\varrho,
\end{equation}
where each $\tilde{\mathcal{L}}_j$ is defined analogously to $\mathcal{L}_j$ but with $\hat{b}_j$ replaced by $\tilde{b}_j$. The Hamiltonian can be expressed as the sum of Hamiltonians for $N-1$ uncoupled and two linearly-coupled oscillators:
\begin{equation}
\hat{H}=\Omega\hat{a}^{\dagger}\hat{a}+\sum_{j=1}^N\omega\tilde{b}_j^\dagger\tilde{b}_j+g\bigl(\hat{a}+\hat{a}^\dagger\bigr)\bigl(\tilde{b}_1+\tilde{b}_1^\dagger\bigr),
\end{equation}
Note that it is only because the thermal baths all have the same temperature that the Liouvillians $\tilde{L}_j$ are diagonal in the new basis. As shown in Appendix I, the adiabatic elimination of $\hat{a}$ and application of the rotating-wave approximation leads to a shift of the oscillator $\tilde b_1$'s frequency and a modification of its coupling to the baths. In the single-oscillator case the adiabatic elimination leads to the appearance of an effective thermal bath. In the new basis, this means that we obtain an effective master equation for the reduced oscillator-only subsystem
\begin{equation}
\label{eq:ReducedME}
\dot\rho=-i[\hat{H}_\mathrm{eff},\rho]+\tilde{\mathcal{L}}^\prime\rho+\sum_{j=2}^N\tilde{\mathcal{L}}_j\rho,
\end{equation}
where
\begin{equation}
\hat{H}_\mathrm{eff}=\omega^\prime\tilde{b}_1^\dagger\tilde{b}_1+\sum_{j=2}^N\omega\tilde{b}_j^\dagger\tilde{b}_j,
\end{equation}
and
\begin{equation}
\tilde{\mathcal{L}}^\prime\rho=\gamma^\prime\bigl[(n^\prime+1)\bigl(2\tilde{b}_1\rho\tilde{b}_1^\dagger-\tilde{b}_1^\dagger\tilde{b}_1\rho-\rho\tilde{b}_1^\dagger\tilde{b}_1\bigr)+n^\prime\bigl(2\tilde{b}_1^\dagger\rho\tilde{b}_1-\tilde{b}_1\tilde{b}_1^\dagger\rho-\rho\tilde{b}_1\tilde{b}_1^\dagger\bigr)\bigr].
\end{equation}
The reduced master equation \eref{eq:ReducedME} has a steady-state solution $\rho_\mathrm{ss}$ given by a tensor product of $N$ thermal states $\rho_j$, with an occupation number given by $n^\prime$ for $j=1$ and $n$ for $j\geq2$. Writing the heat flow in the normal-mode basis yields
\begin{equation}
J_l=2\omega\gamma\Biggl(n-\sum_{j,k=1}^N\matu{G}_{jl}\matu{G}_{kl}\langle\tilde{b}_j^\dagger\tilde{b}_k\rangle\Biggr).
\end{equation}
Because of the uncorrelated nature of the steady state, we can immediately write that $\langle\tilde{b}_j^\dagger\tilde{b}_k\rangle=0$ if $j\neq k$, i.e.,
\begin{equation}
J_l=2\omega\gamma\sum_{j=1}^N\matu{G}_{jl}^2\bigl(n-\langle\tilde{b}_j^\dagger\tilde{b}_j\rangle\bigr),
\end{equation}
where we have also taken the sum out of the parentheses by exploiting the properties of the orthonormal matrix $\matu{G}$. By the very nature of the steady state, however,
\begin{equation}
\langle\tilde{b}_j^\dagger\tilde{b}_j\rangle=n+(n^\prime-n)\delta_{j,1},
\end{equation}
so that
\begin{equation}
J_l=2\omega\gamma\matu{G}_{1l}^2(n-n^\prime)=2\omega\gamma(n-n^\prime)g_l^2/g^2.
\end{equation}
Therefore, the heat flowing into or out of the mechanical subsystem is simply
\begin{equation}
J_\mathrm{m}:=\sum_{l=1}^NJ_l=2\omega\gamma(n-n^\prime),
\end{equation}
which is nonzero because we have traced out the $(N+1)$\textsuperscript{st} oscillator. Let us note that $n-n^\prime\to0$ when $g\to0$. The heat flowing into or out of this oscillator must therefore be
\begin{equation}
J_\mathrm{c}:=-\sum_{l=1}^NJ_l=2\omega\gamma(n^\prime-n),
\end{equation}
to maintain balance, i.e., $J_\mathrm{m}+J_\mathrm{c}=0$.

\section{Application to optomechanical arrays}
\subsection{Two-element optomechanical array}\label{sec:OM}
As an application, we consider a system composed of two mechanical oscillators in which each identical, independent oscillator is dispersively coupled by radiation pressure to the same cavity field mode. The mechanical oscillators have identical frequency $\omega$ and equal damping rate $\gamma$ into two independent Markovian thermal baths held at possibly different temperatures, yielding mean thermal occupation numbers $n_1$ and $n_2$ for the mechanical modes in absence of the field. We assume operation in the linearised regime for the optomechanical interaction~\cite{Aspelmeyer2013} in which the number of intracavity photons is large. Without loss of generality we also consider a situation in which the cavity field couples to the centre-of-mass motion of the pair of mechanical oscillators, leaving the relative mode of motion uncoupled.\footnote{Note that the opposite situation can be realised in, e.g., a ``transmissive'' configuration~\cite{Xuereb2012} or in the double-cavity geometry of Ref.~\cite{Pinard2005}.} Under these conditions the Hamiltonian and Liouvillian of the system are given by
\begin{eqnarray}
H&=-\Delta\hat{a}^{\dagger}\hat{a}+\omega(\hat{b}_1^{\dagger}\hat{b}_1+\hat{b}_2^{\dagger}\hat{b}_2)+g(\hat{a}+\hat{a}^{\dagger})\bigl[(\hat{b}_1+\hat{b}_1^{\dagger})+(\hat{b}_2+\hat{b}_2^{\dagger})\bigr]/\sqrt{2},\\
\mathcal{L}&=\mathcal{L}_1+\mathcal{L}_2+\mathcal{L}_\mathrm{a}.
\end{eqnarray}
We have defined the cavity field detuning $\Delta=\Omega_\mathrm{L}-\Omega$, where $\Omega_\mathrm{L}$ is the frequency of the driving field, assumed to be monochromatic, and have transformed our system to a frame rotating at the frequency $\Omega_\mathrm{L}$. The enhanced optomechanical coupling rate $g$ and the Liouvillians $\mathcal{L}_{1(2)}$ and $\mathcal{L}_\mathrm{a}$ are defined as in Eqs.~\eref{eq:Lj} and~\eref{eq:La}, respectively. Introducing the relative and centre-of-mass modes
\begin{equation}
\hat{b}_\mathrm{r}=(\hat{b}_1-\hat{b}_2)/\sqrt{2},\qquad \hat{b}_\mathrm{c}=(\hat{b}_1+\hat{b}_2)/\sqrt{2},
\end{equation}
these can be recast as
\begin{eqnarray}
H&=-\Delta\hat{a}^{\dagger}\hat{a}+\omega(\hat{b}_\mathrm{r}^{\dagger}\hat{b}_\mathrm{r}+\hat{b}_\mathrm{c}^{\dagger}\hat{b}_\mathrm{c})+g(\hat{a}+\hat{a}^{\dagger})(\hat{b}_\mathrm{c}+\hat{b}_\mathrm{c}^{\dagger}),\\
\mathcal{L}&=\mathcal{L}_\mathrm{r}+\mathcal{L}_\mathrm{c}+\mathcal{L}_\mathrm{r,c}+\mathcal{L}_\mathrm{a},
\end{eqnarray}
with $\mathcal{L}_\mathrm{r,c}$ a Liouvillian including the correlations between the rotated baths.

Adiabatically eliminating the cavity field in the weak optomechanical coupling regime leads to an effective mechanical frequency $\omega^\prime=\omega+\Lambda$, due to the optical spring effect, as well as an effective damping rate $\gamma^\prime=\gamma+\bar{\gamma}$, for the centre-of-mass mode (see Appendix I), where
\begin{eqnarray}
\label{eq:Lambda}\Lambda&=\frac{2g^2\Delta(\Delta^2-\omega^2+\kappa^2)}{[(\Delta+\omega)^2+\kappa^2][(\Delta-\omega)^2+\kappa^2]},\\
\label{eq:gammabar}\bar{\gamma}&=\beta_+-\beta_-=\frac{G^2\kappa}{(\Delta+\omega)^2+\kappa^2}-\frac{G^2\kappa}{(\Delta-\omega)^2+\kappa^2}.
\end{eqnarray}
The field fluctuations also give rise to an additional effective bath, coupling to the centre-of-mass mode only, whose Liouvillian has the form
\begin{equation}
\bar{\mathcal{L}}=\beta_+D[\hat{b}_\mathrm{c}]+\beta_-D[\hat{b}_\mathrm{c}^{\dagger}]
\end{equation}
Assuming a red-detuned cavity field ($\Delta<0$) for which $\beta_+>\beta_-$, this term describes the coupling with a thermal bath with coupling rate and occupation number defined by
\begin{equation}
\beta_+=\bar{\gamma}(\bar{n}+1),\quad\beta_-=\bar{\gamma}\bar{n}\hspace{0.5cm}
\Rightarrow\hspace{0.5cm}\bar{\gamma}=\beta_+-\beta_-,\quad\bar{n}=\frac{\beta_-}{\beta_+-\beta_-}
\end{equation}
The effective Hamiltonian and Liouvillian after adiabatic elimination then read
\begin{eqnarray}
H_{\mathrm{eff}}&=(\omega+\Lambda)\hat{b}_\mathrm{c}^{\dagger}\hat{b}_\mathrm{c}+\omega \hat{b}_\mathrm{r}^{\dagger}\hat{b}_\mathrm{r},\\
\mathcal{L}_{\mathrm{eff}}&=\mathcal{L}_\mathrm{r}+\mathcal{L}_\mathrm{c}+\mathcal{L}_\mathrm{r,c}+\bar{\mathcal{L}},
\end{eqnarray}
or, going back to the bare mechanical basis,
\begin{eqnarray}
H_{\mathrm{eff}}&=(\omega+\Lambda/2)(\hat{b}_1^{\dagger}\hat{b}_1+\hat{b}_2^{\dagger}\hat{b}_2)+(\Lambda/2)(\hat{b}_1^{\dagger}\hat{b}_2+\hat{b}_1\hat{b}_2^{\dagger}),\\
\mathcal{L}_{\mathrm{eff}}&=\mathcal{L}_1+\mathcal{L}_2+\bar{\mathcal{L}}.
\end{eqnarray}
In the regime of interest, $\Lambda\ll\omega$, so that we may approximate
\begin{equation}
H_{\mathrm{eff}}=\omega(\hat{b}_1^{\dagger}\hat{b}_1+\hat{b}_2^{\dagger}\hat{b}_2)+(\Lambda/2)(\hat{b}_1^{\dagger}\hat{b}_2+\hat{b}_1\hat{b}_2^{\dagger}).
\end{equation}
These indeed describe the dynamics of a pair of oscillators, mutually coupled with a strength $\Lambda$ and in contact with two independent baths and a common bath described by the Liouvillians $\mathcal{L}_{1,2}$ and $\bar{\mathcal{L}}$, respectively. The master equation that emerges from this model is a special case of a more general master equation; in the next section we shall describe this more general situation and derive the steady state of the system.

\subsection{Two oscillators with independent and common thermal baths}\label{sec:OMIndependentCB}
A general treatment of the dynamics of a pair of coupled oscillators in contact with either independent baths or a common bath can be found in Ref.~\cite{Galve2010}. Motivated by the results of the previous section, we consider the same Hamiltonian as in Ref.~\cite{Galve2010}:
\begin{equation}
\hat{H}_\mathrm{eff}=\frac{\omega}{2}\sum_{j=1}^N\bigl(\hat{p}_j^2+\hat{x}_j^2\bigr)+\Lambda\hat{x}_1\hat{x}_2,
\end{equation}
but focus on the case of two identical oscillators (equal mass and frequency) and each in simultaneous contact with both an independent and a common bath; we have defined $\hat{x}_j=(\hat{b}_j+\hat{b}_j^\dagger)/\sqrt{2}$ and $\hat{p}_j=(\hat{b}_j-\hat{b}_j^\dagger)/(i\sqrt{2})$ ($j=1,2$). We assume the baths to be Markovian, the independent bath temperatures being given by $n_j$, and the common bath temperature by $\bar{n}$. In order to obtain an effective master equation in the limit $\gamma,\bar{\gamma}\ll\Lambda\ll\omega$, we employ these assumptions to follow the procedure of Ref.~\cite{Galve2010}, which we will not detail explicitly but finally yields
\begin{eqnarray}
\label{eq:rhoeff2osc}
\nonumber\dot{\rho}=-i[\hat{H}_\mathrm{eff},\rho]&-\gamma\sum_{j=1}^2\Bigl(i[\hat{x}_j,\{\hat{p}_j,\rho\}]+(2n_j+1)[\hat{x}_j,[\hat{x}_j,\rho]]\Bigr)\\
&-\frac{\bar{\gamma}}{2}\sum_{j=1}^2\Bigl(i[\hat{x}_+,\{\hat{p}_j,\rho\}]+(2\bar{n}+1)[\hat{x}_+,[\hat{x}_j,\rho]]\Bigr),
\end{eqnarray}
where $\hat{x}_+=\hat{x}_1+\hat{x}_2$. Under the rotating-wave approximation, where terms of the form $\hat{b}_1\hat{b}_2$ and $\hat{b}_1^{\dagger}\hat{b}_2^{\dagger}$ are neglected based on the approximation that their effects average out over the longer time-scales of relevance to the problem, the model described by this master equation reduces to the optomechanical model derived in Sec.~\ref{sec:OM}. In the remainder of this section we show that, under the conditions outlined below, we can solve this more general model explicitly and apply this solution to the situation in Sec.~\ref{sec:OM}.

From this master equation, and exploiting the standard commutation relations, we derive a closed system of equations describing the temporal evolution of the sixteen second-order moments, which is shown in Appendix II. Under the assumption that the initial state is Gaussian, the dynamics described by these equations preserves the Gaussian nature of the state at all time. From these equations it is straightforward to compute the steady state occupation numbers of both oscillators,
\begin{equation}
n_j'=\expt{\hat{b}_j^\dagger\hat{b}_j}=\bigl[\expt{\hat{x}_j^2}+\expt{\hat{p}_j^2}-1\bigr]/2,
\end{equation}
and thereby calculate the heat flow in this system from \eref{eq:JlBare}. One gets
\begin{eqnarray}
\label{eq:SteadyStateOcc}
\nonumber n_j'=&\frac{2\gamma n_j+\bar{\gamma}\bar{n}}{2\gamma+\bar{\gamma}}+\frac{\bar{\gamma}^2}{2(2\gamma+\bar{\gamma})(\gamma+\bar{\gamma})}\left(\frac{n_1+n_2}{2}-\bar{n}\right)\\
&+(-1)^j\frac{2\gamma\Lambda^2}{(2\gamma+\bar{\gamma})[(2\gamma+\bar{\gamma})^2+\Lambda^2]}\frac{n_1-n_2}{2}\qquad(j=1,2),
\end{eqnarray}
which leads to the following expression for the steady-state heat flows
\begin{eqnarray}
\label{eq:SteadyStateHF}
\nonumber J_j=4\omega\gamma\Biggl\{&\frac{\bar{\gamma}}{2\gamma+\bar{\gamma}}(n_j-\bar{n})+\frac{\bar{\gamma}^2}{2(2\gamma+\bar{\gamma})(\gamma+\bar{\gamma})}\Biggl(\bar{n}-\frac{n_1+n_2}{2}\Biggr)\\
&-(-1)^j\frac{\gamma\Lambda^2}{(2\gamma+\bar{\gamma})[(2\gamma+\bar{\gamma})^2+\Lambda^2]}(n_1-n_2)\Biggr\}\quad (j=1,2).
\end{eqnarray}
These expressions bring together the work of Ref.~\cite{Galve2010} and our generic expression for the heat flow, Eq.~\eref{eq:JlBare}. It expresses the heat flow through an array of two oscillators that are coupled not only to independent baths but also to a third, ``correlated'' or common, bath, which sets up an effective interaction between the two oscillators, altering the nature of the heat transport. In the next section, we discuss this interplay between the independent and common baths in more detail.

\subsection{Discussion}\label{sec:Discussion}
\subsubsection{Two-oscillator case.}
It is interesting to look at these expressions in various regimes of interest. The regime of large mutual coupling between the oscillators corresponds to the large optical spring regime, which can be achieved in the bad-cavity limit of optomechanics, $\kappa\gg\omega$. In this regime, and for $\Delta\sim -\kappa$, Eqs.~\eref{eq:Lambda} and~\eref{eq:gammabar} give $|\Lambda|\sim g^2/\kappa\gg \bar{\gamma}\sim2g^2\omega/\kappa^2$. In contrast, the regime of large coupling to the effective common bath can be achieved in the resolved sideband regime of optomechanics, $\omega\gg\kappa$. For $\Delta=-\omega$, one obtains $|\Lambda|\sim g^2/(2\omega)\ll \bar{\gamma}\sim g^2/\kappa$.
\begin{enumerate}
\item In the large coupling regime $\Lambda\gg\gamma,\bar{\gamma}$, one has
\begin{equation}
n_j'=\frac{2\gamma+\bar{\gamma}}{2(\gamma+\bar{\gamma})}\frac{n_1+n_2}{2}+\frac{\bar{\gamma}}{2(\gamma+\bar{\gamma})}\bar{n}\qquad(j=1,2)
\end{equation}
which we can understand in two different limits. (a)~When the coupling to the independent baths is larger than that to the common bath ($\gamma\gg\bar{\gamma}$), the mean independent bath temperature $(n_1+n_2)/2$. (b)~When the damping into the common bath dominates over the mutual coupling, the whole system equilibrates at the mean of the common and independent bath temperatures
\begin{equation}
n_j'=\frac{\bar{n}}{2}+\frac{n_1+n_2}{4}\qquad(j=1,2)
\end{equation}
\item If the independent baths are held at the same temperature $n$, the term due to the mutual coupling vanishes and one gets
\begin{equation}\label{eq:sametemperature}
n_j'=\frac{2\gamma+\bar{\gamma}}{2(\gamma+\bar{\gamma})}n+\frac{\bar{\gamma}}{2(\gamma+\bar{\gamma})}\bar{n},\quad J_j=\omega\gamma\frac{\bar{\gamma}}{\gamma+\bar{\gamma}}(n-\bar{n})\quad(j=1,2)
\end{equation}
This is consistent with the limit of radiation pressure cooling in optomechanics~\cite{Wilsonrae2007, Marquardt2007, Genes2008, Dantan2008}, which predicts that the (centre-of-mass) mode coupled to the field is cooled down to $\bar{n}$, $\bar{\gamma}\gg\gamma$. Since the uncoupled (relative motion) mode's occupancy remains $n$, this means that, in the bare basis, $n_j'\rightarrow(n+\bar{n})/2$ when $\bar{\gamma}\gg\gamma$.
\end{enumerate}

\begin{figure}
\centering
\includegraphics[width=0.33\textwidth]{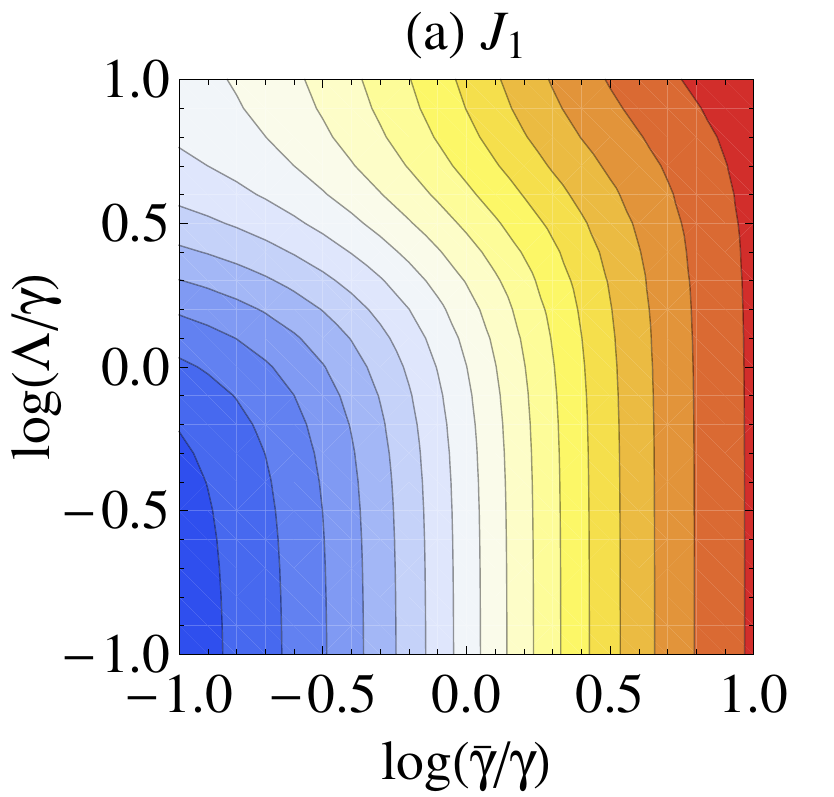}\includegraphics[width=0.33\textwidth]{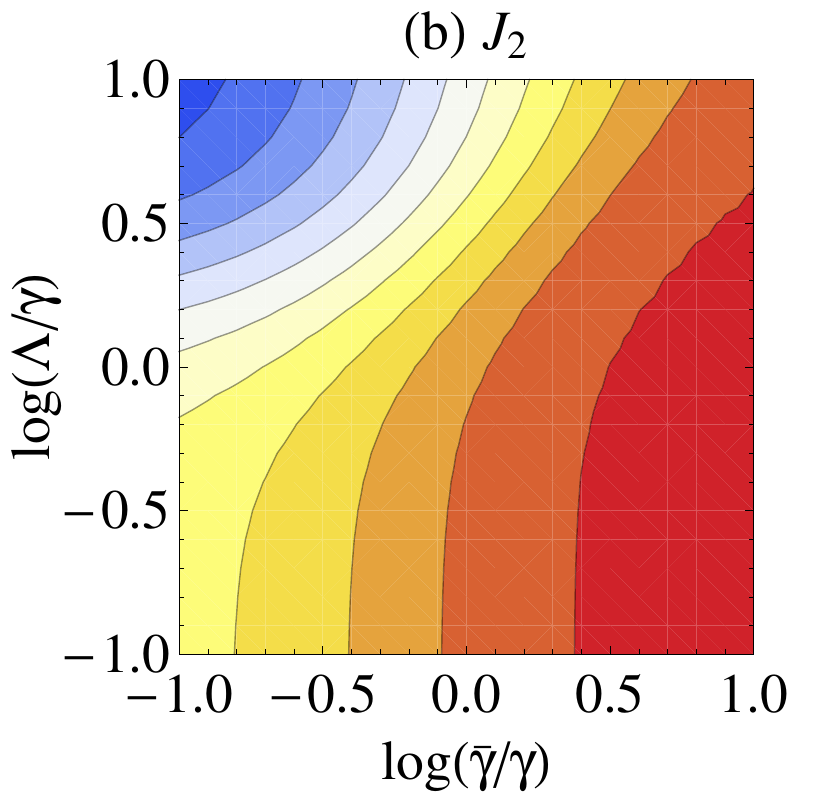}\includegraphics[width=0.33\textwidth]{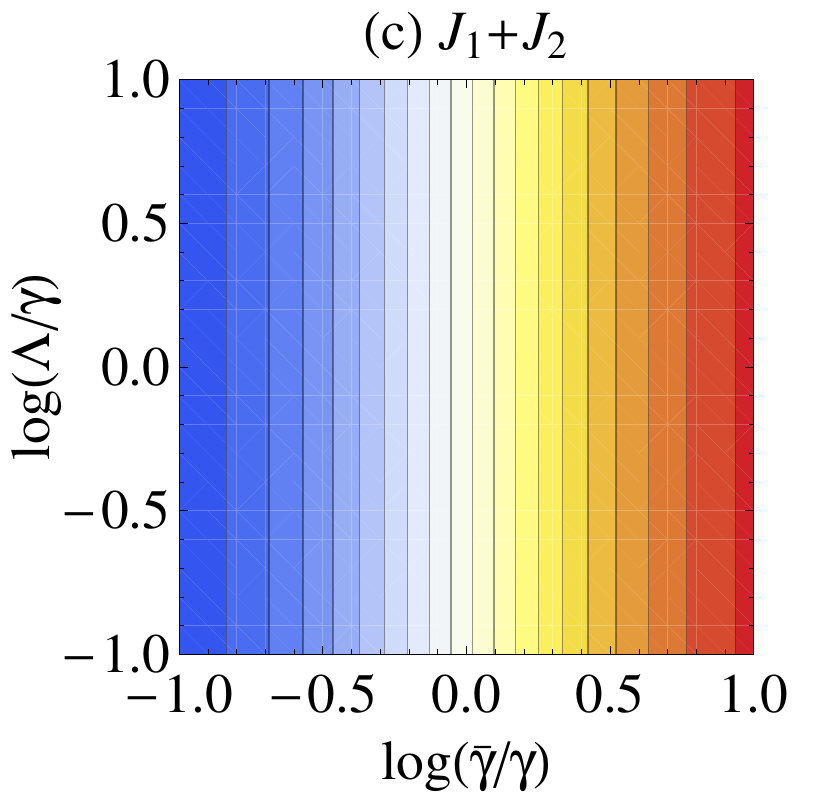}\\
\includegraphics[width=0.33\textwidth]{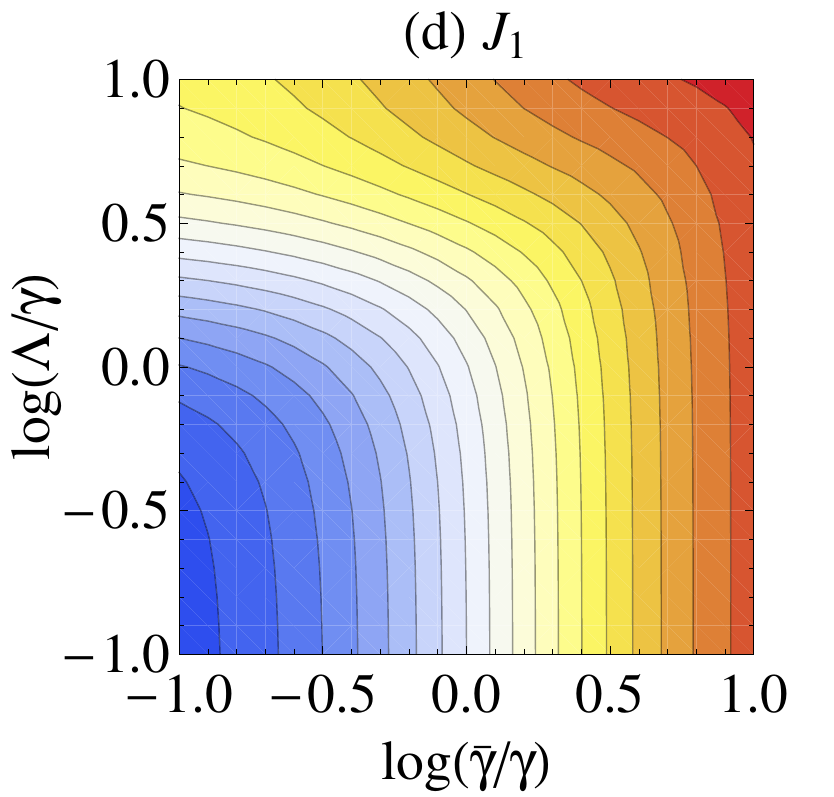}\includegraphics[width=0.33\textwidth]{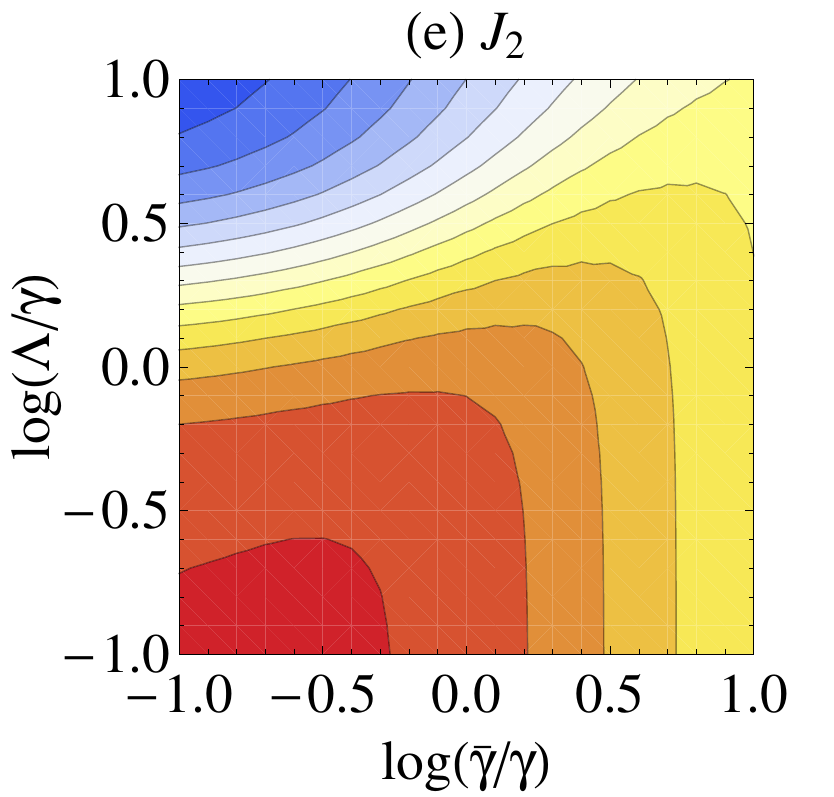}\includegraphics[width=0.33\textwidth]{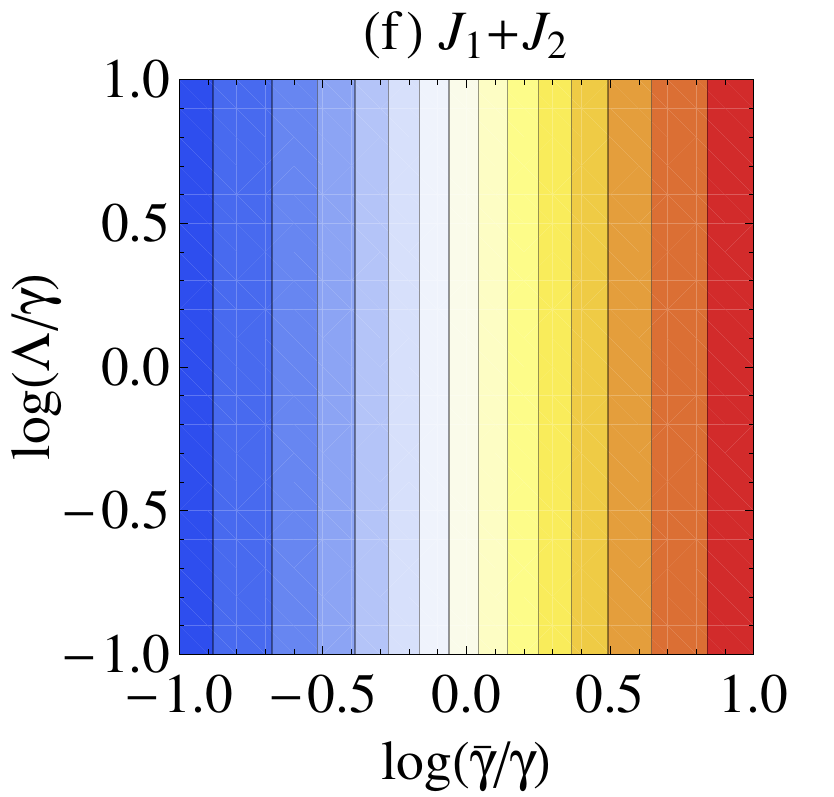}\\
\caption{Heat flows $J_1$ (a,d), $J_2$ (b,e), $J_1+J_2$ (c,f) (in arbitrary units) as a function of $\Lambda/\gamma$ and $\bar{\gamma}/\gamma$, for the two-oscillator case. Figures (a,b,c) and (d,e,f) correspond to a small ($n_1=2n_2$) and a large ($n_1=10n_2$) temperature difference between the independent baths, respectively. The common bath has $n_0=0$ in both cases.} \label{figN2}
\end{figure}
As an illustration, Fig.~\ref{figN2} shows the individual heat flows $J_1$ and $J_2$, as well as the total heat flow through the mechanics $J_1+J_2$, as a function of $\Lambda/\gamma$ and $\bar{\gamma}/\gamma$, in the case where the first oscillator is coupled to a higher temperature bath than the second, and for a common bath at zero temperature. The heat flow from the hotter oscillator increases both with $\Lambda$ and $\bar{\gamma}$, as both the common bath and the other oscillator's bath have a lower temperature. The heat flow from the colder oscillator, however, becomes negative for a large mutual coupling, as this coupling tends to equalise the temperature of both oscillators. Moreover, depending on the temperature difference between the independent baths, the heat flow from the cold oscillator can be seen to either increase [Fig.~\ref{figN2}(b)] or decrease [Fig.~\ref{figN2}(e)] with the coupling with the common bath, as the cold oscillator gets either cooled or heated by the combined action of the cold common bath and the other hot oscillator. The total heat flow through the mechanics is in contrast independent of the mutual coupling and steadily increases with the coupling to the common bath, as the optical field globally takes away heat from the mechanics.

\subsubsection{$N$-oscillator case}
\begin{figure}
\centering
\includegraphics[width=0.47\textwidth]{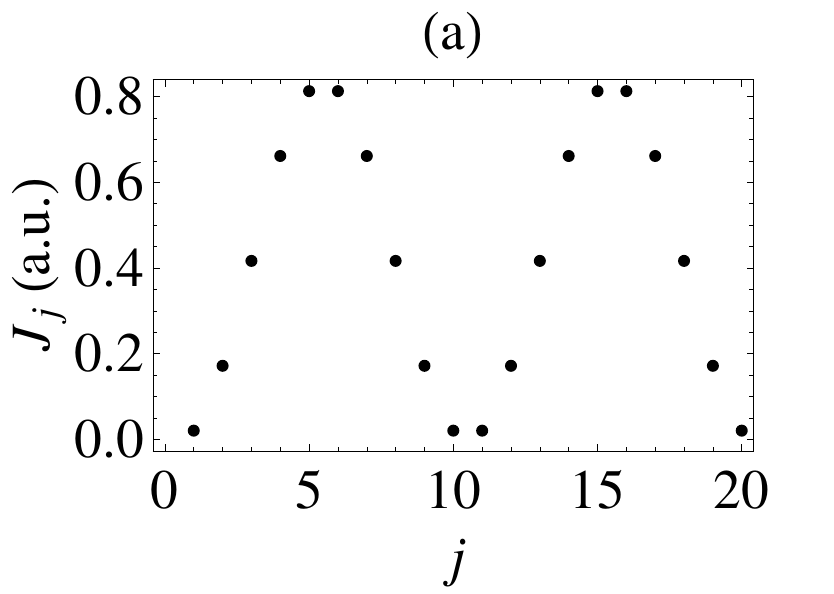}\includegraphics[width=0.5\textwidth]{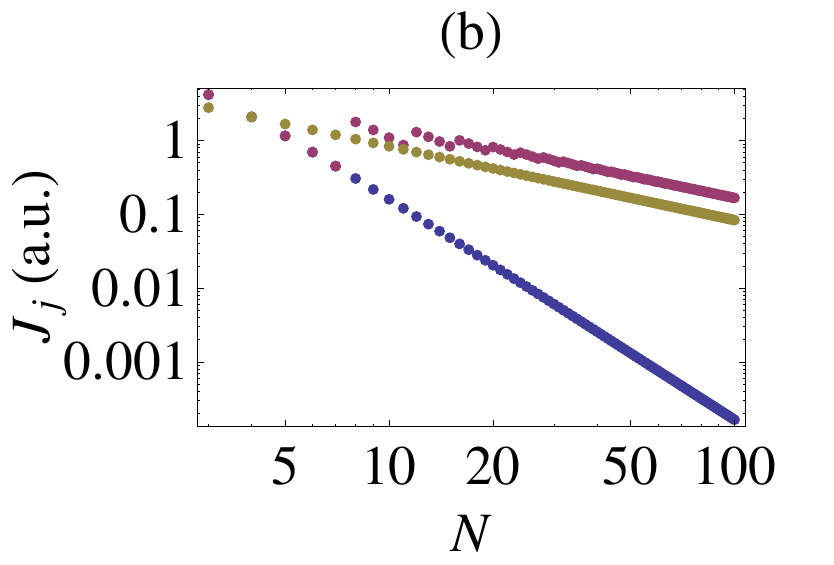}
\caption{(a) Heat flow $J_j$, in arbitrary units, for a 20-element array with the couplings given by Eq.~\eref{eq:gj}. (b) Heat flow at the beginning ($j=1$, blue) and at one-quarter ($j=\left \lfloor{N/4}\right \rfloor$, magenta) of the array, as a function of $N$. The averaged heat flow per element $\bar{J}$ (mustard) is also shown.} \label{figfourier}
\end{figure}

In the case of $N$ baths all at \emph{equal} temperature, one can use this formalism together with the ideas developed in Sec.~\ref{sec:EqualTemp} to find
\begin{equation}
n_j'=n+\frac{g_j^2}{g^2}\frac{\bar{\gamma}}{\gamma+\bar{\gamma}}(\bar{n}-n),\quad J_j=2\omega\gamma\frac{g_j^2}{g^2}\frac{\bar{\gamma}}{\gamma+\bar{\gamma}}(n-\bar{n})\quad(j=1-N)
\label{n_j_eq}
\end{equation}
This means that the heat flow through the $j$\textsuperscript{th} mechanical element is proportional to the temperature difference between the independent thermal baths and the field bath, weighted by the branching ratio of the damping rates $\eta=\bar{\gamma}/(\gamma+\bar{\gamma})$ and the relative optomechanical coupling strength of the $j$\textsuperscript{th} element to the field. Note that, because of the normalisation of the $g_j$, the total heat flow through the array is independent of the system size\footnote{We remark here that any intrinsic dependence of $g$ itself on the system size can be counteracted by varying the driving strength appropriately.}:
\begin{equation}
J_\mathrm{m}=2\omega\gamma\eta(n-\bar{n})=2\omega\gamma(n-n')
\label{total_curr:eq}
\end{equation}
where $n'=(1-\eta)n+\eta \bar{n}$ is the final occupation number of the collective mode $\tilde{b}_1$. Let us also note that, while the {\it averaged} heat flow per element, $\bar{J}:=\frac{1}{N}\sum_jJ_j=J_\mathrm{m}/N$, scales as the inverse of the length of the arrays, as expected from the Fourier law~\cite{Dhar2008}, the {\it local} heat flow depends on the form of the individual optomechanical coupling. To take an example, for a field whose wavelength is chosen such that the whole array is ``transmissive,'' the $g_j$ can be shown to have a sinusoidal dependence with the element position in the array~\cite{Xuereb2012,Xuereb2013}
\begin{equation}
\label{eq:gj}
g_j=g \sqrt{\frac{2}{N}}\,\sin\Biggl(2 \pi \frac{j-1/2}{N}\Biggr)\quad(N>2).
\label{gi:eq}
\end{equation}
One sees from Eq.~\eref{n_j_eq} that the currents flowing through the different oscillators have quite different behaviours in the large-$N$ limit. Indeed, as shown in Fig.~\ref{figfourier}, at the extremities (or the centre) of the array, one has $J_{j}\sim1/N^3$ as $N\rightarrow \infty$, while for $j\sim N/4$ or $3N/4$, one obtains $J_{j}\sim1/N$. Choosing the form of the optomechanical couplings thus offers some freedom in tuning the heat flow through individual elements.

\subsubsection{Experimental implementation}
The generic features discussed previously could in principle be observed in a broad range of optomechanical systems: In the optical domain, these could be arrays of flexible membranes~\cite{Thompson2008}, toroidal cavities with indentations~\cite{Arbadi2011}, optomechanical crystals~\cite{Eichenfield2009}, or ensembles of cold atoms in optical cavities~\cite{Brahms2012,Botter2013}; in the microwave domain, micromechanical elements coupled to superconducting microwave resonator fields~\cite{Teufel2011,Massel2012} are one such possibility. Depending on the system considered, the measurement of the oscillators' thermal occupancy can be performed optically via sideband thermometry~\cite{Chan2011,Purdy2014} or collective mode readout~\cite{Brahms2012,Xuereb2012}, or electrically by functionalizing the elements~\cite{Adiga2013,Bagci2013,Andrews2014} or by coupling the mechanical elements to additional microwave resonators or to artificial atoms, as demonstrated in, e.g., Refs.~\cite{OConnell2010,Pirkkalainen2013}. The two-oscillator scenario could be for instance realised using two micromirrors~\cite{Arcizet2006,Gigan2006} in the double-cavity geometry of \cite{Pinard2005} in which the motion of the centre-of-mass and relative-motion modes can be addressed and readout by two optical fields appropriately detuned from the cavity resonance. Alternatively, one could use a pair of partially transmitting, flexible membranes positioned in an optical cavity driven by optical fields with specific wavelengths~\cite{Xuereb2012}.

\section{Conclusion}\label{sec:Conclusion}
We have investigated the transport of heat in a system of $N$ quantum oscillators coupled to common and independent baths and derived analytical expressions for the steady state occupancy and heat flow. The obtained results are, among others, relevant in the context of optomechanical arrays where by choosing the form of the coupling between the optical field and the mechanics one can engineer effective couplings and baths for the mechanical oscillators, and thereby tune the heat flow through individual elements. While the present work focussed on the situation of an optical field in a coherent state---a common oscillator coupled to a zero-temperature bath---the same approach could be used to tackle the case of optomechanical interactions with (Gaussian) fields exhibiting nonclassical correlations, such as squeezing~\cite{Zhang2003, Pinard2005}.

\section*{Acknowledgements}
This work was supported by the COST Action MP1209 ``Thermodynamics in the Quantum Regime,'' the Royal Commission for the Exhibition of 1851, and the Danish Council for Independent Research (Sapere Aude program).

\section*{Appendix I: Adiabatic elimination of the optical mode from the master equation}
The purpose of this appendix is to introduce to the unfamiliar reader the main steps involved in the adiabatic elimination of a fast system, in our case an optical field described by the annihilation operator $\hat{a}$, from a coupled system involving the optical field and a mechanical mode, described by the annihilation operator $\hat{b}$. Our treatment can easily be generalised to many mechanical modes and follows very closely the exposition in the Supplemental Information of Ref.~\cite{Xuereb2014}. We refer the interested reader to alternative expositions, e.g., in Ref.~\cite{Gardiner2004}, for further detail. To start, we divide the total Hamiltonian of the system into three components. First is the free Hamiltonian
\begin{equation}
\hat{H}_\mathrm{free}=-\Delta\hat{a}^\dagger\hat{a}+\omega\hat{b}^\dagger\hat{b}^{\phantom{\dagger}},
\end{equation}
which we write in the rotating from of the driving field (introduced below), such that $\Delta:=\Omega_\mathrm{L}-\Omega$ is the difference between the resonance frequency of mode `a' and its driving field. Next, we must consider the Hamiltonian that couples the two subsystems,
\begin{equation}
\hat{H}_\mathrm{int}=-g\bigl(\hat{b}+\hat{b}^\dagger\bigr)\hat{a}^\dagger\hat{a}.
\end{equation}
Finally, we add the Hamiltonian that acts to drive the optical field to a non-zero mean value,
\begin{equation}
\hat{H}_\mathrm{d}=i\sqrt{2\kappa}\alpha_\mathrm{in}\bigl(\hat{a}^\dagger-\hat{a}\bigr);
\end{equation}
here, the (coherent) driving field amplitude $\alpha_\mathrm{in}$ is taken to be real for simplicity. The compound system is also acted on by dissipation, which we concisely model using the superoperators
\begin{equation}
D[\hat{c}]\varrho:=2\hat{c}\varrho\hat{c}^\dagger-\hat{c}^\dagger\hat{c}\varrho-\varrho\hat{c}^\dagger\hat{c},
\end{equation}
and
\begin{equation}
D_\mathrm{th}[\hat{c}]:=\bigl(n+1\bigr)D[\hat{c}]+nD[\hat{c}^\dagger]\,,
\end{equation}
where $n$ gives the mean number of excitations present in the bath that $\hat{c}$ is coupled to. We now label the (amplitude) decay rates of $\hat{a}$ and $\hat{b}$ by $\kappa$ and $\gamma$, respectively, such that the full master equation of the system takes the form
\begin{equation}
\dot{\varrho}=-i\bigl[\hat{H}_\mathrm{free}+\hat{H}_\mathrm{int}+\hat{H}_\mathrm{d},\varrho\bigr]+\kappa D[\hat{a}]\varrho+\gamma D_\mathrm{th}\!\bigl[\hat{b}\bigr]\varrho\,.
\end{equation}
We now proceed by defining two simplifying unitary transformations; we shift both $\hat{a}$ and $\hat{b}$ by an as-yet undetermined complex number:
\begin{equation}
\hat{U}_\mathrm{a}^{\phantom{\dagger}}\,\hat{a}\,\hat{U}_\mathrm{a}^\dagger=\hat{a}+\alpha\,\mathrm{\ and\ }\hat{U}_\mathrm{m}^{\phantom{\dagger}}\,\hat{b}\,\hat{U}_\mathrm{m}^\dagger=\hat{b}+\beta\,.
\end{equation}
We determine $\alpha$ and $\beta$ by demanding that they act to cancel out the contribution of $\hat{H}_\mathrm{d}$ in the master equation, that $\hat{a}$ and $\hat{b}$ are shifted to have zero mean value. Furthermore, we suppose that $\left|\alpha\right|\gg1$, such that we can neglect the term in the transformed $\hat{H}_\mathrm{int}$ that is proportional to $\hat{a}^\dagger\hat{a}$. Finally, we obtain a transformed Hamiltonian that is quadratic in the operators:
\begin{equation}
\hat{H}_\mathrm{trans}:=-\Delta\hat{a}^\dagger\hat{a}+\omega\hat{b}^\dagger\hat{b}-g\bigl(\hat{b}+\hat{b}^\dagger\bigr)\bigl(\alpha^\ast\hat{a}+\alpha\hat{a}^\dagger\bigr).
\end{equation}
We have further redefined $\Delta$ in order to absorb a shift in the cavity frequency introduced by $\beta$. It is now convenient to consider the three terms making up $\hat{H}_\mathrm{trans}$ separately. In order to do this, we define the system $\hat{H}_\mathrm{s}:=\omega\hat{b}^\dagger\hat{b}$, bath $\hat{H}_\mathrm{b}:=-\Delta\hat{a}^\dagger\hat{a}$, and interaction $\hat{H}_\mathrm{i}:=-g\bigl(\hat{b}+\hat{b}^\dagger\bigr)\bigl(\alpha^\ast\hat{a}+\alpha\hat{a}^\dagger\bigr)$ Hamiltonians. We also define the associated Liouvillians
\begin{eqnarray}
\mathcal{L}_\mathrm{s}\varrho&:=-i\bigl[\hat{H}_\mathrm{s},\varrho\bigr]+\gamma D_\mathrm{th}\!\bigl[\hat{b}\bigr]\varrho,\\
\mathcal{L}_\mathrm{b}\varrho&:=-i\bigl[\hat{H}_\mathrm{b},\varrho\bigr]+\gamma D[\hat{a}]\varrho,
\end{eqnarray}
and
\begin{equation}
\mathcal{L}_\mathrm{i}\varrho:=-i\bigl[\hat{H}_\mathrm{i},\varrho\bigr],
\end{equation}
respectively. With this notation, the transformed master equation acquires the simple form $\dot{\varrho}=\bigl(\mathcal{L}_\mathrm{s}+\mathcal{L}_\mathrm{b}+\mathcal{L}_\mathrm{i}\bigr)\varrho$. Our goal in this section is to derive an effective equation of motion for the density matrix describing the mechanical subsystem, with mode `a' eliminated from it. In common with other such eliminations, our results will only be valid to lowest order in the coupling strength $g$.

The approach we follow uses projection operators to effectively project the master equation and achieve this elimination. The first projection operator we require takes the form
\begin{equation}
\mathbb{P}\varrho:=\rho_\mathrm{a}\otimes\Tr_\mathrm{a}\varrho\,,
\end{equation}
where $\rho_\mathrm{a}$ is the vacuum state for mode `a' and corresponds to the solution of the master equation for mode `a' alone. We define $\mathbb{I}$ as the identity operator and a further projection operator $\mathbb{Q}=\mathbb{I}-\mathbb{P}$. Following~\cite[\S5.2.1]{Gardiner2004} we note that
\begin{eqnarray}
\mathbb{P}\mathcal{L}_\mathrm{i}\mathbb{P}=0,\mathrm{\ and}\\
\mathbb{Q}\bigl(\mathcal{L}_\mathrm{s}+\mathcal{L}_\mathrm{b}\bigr)=\bigl(\mathcal{L}_\mathrm{s}+\mathcal{L}_\mathrm{b}\bigr)\mathbb{Q}.
\end{eqnarray}
These projection operators are used to project the full master equation, resulting in the two differential equations
\begin{eqnarray}
\mathbb{P}\dot{\varrho}&=\mathbb{P}\bigl(\mathcal{L}_\mathrm{s}+\mathcal{L}_\mathrm{b}\bigr)\mathbb{P}\varrho+\mathbb{P}\mathcal{L}_\mathrm{i}\mathbb{Q}\varrho,\mathrm{\ and}\\
\mathbb{Q}\dot{\varrho}&=\mathbb{Q}\bigl(\mathcal{L}_\mathrm{s}+\mathcal{L}_\mathrm{b}+\mathcal{L}_\mathrm{i}\bigr)\mathbb{Q}\varrho+\mathbb{Q}\mathcal{L}_\mathrm{i}\mathbb{P}\varrho.
\end{eqnarray}
We now invoke the weak-coupling approximation, assuming $g$ to be small. We can then formally write
\begin{equation}
\mathbb{Q}\varrho(t)=e^{\mathbb{Q}(\mathcal{L}_\mathrm{s}+\mathcal{L}_\mathrm{b}+\mathcal{L}_\mathrm{i})(t-t_0)}\mathbb{Q}\varrho(t_0)+\int_{t_0}^t\rmd\tau\,e^{\mathbb{Q}(\mathcal{L}_\mathrm{s}+\mathcal{L}_\mathrm{b}+\mathcal{L}_\mathrm{i})(t-\tau)}\mathbb{Q}\mathcal{L}_\mathrm{i}\mathbb{P}\varrho(\tau)\,.
\end{equation}
Next, we substitute the lowest-order approximation
\begin{equation}
\mathbb{P}\varrho(\tau)=e^{\mathbb{P}(\mathcal{L}_\mathrm{s}+\mathcal{L}_\mathrm{b})(\tau-t)}\mathbb{P}\varrho(t)
\end{equation}
in the integrand, and we take the initial time $t_0\to-\infty$. Because of our lowest-order-in-$g$ approximation, we ignore the $\mathcal{L}_\mathrm{i}$ in the exponent, yielding
\begin{eqnarray}
\mathbb{P}\dot{\varrho}&\approx\mathbb{P}\bigl(\mathcal{L}_\mathrm{s}+\mathcal{L}_\mathrm{b}\bigr)\mathbb{P}\varrho\nonumber\\
&\quad+\int_0^\infty\rmd\tau\,\mathbb{P}\mathcal{L}_\mathrm{i}e^{\mathbb{Q}(\mathcal{L}_\mathrm{s}+\mathcal{L}_\mathrm{b})\tau}\mathbb{Q}\mathcal{L}_\mathrm{i}e^{-\mathbb{P}(\mathcal{L}_\mathrm{s}+\mathcal{L}_\mathrm{b})\tau}\mathbb{P}\varrho(t).
\end{eqnarray}
Next, we notice that $[\mathbb{Q},\mathcal{L}_\mathrm{s}+\mathcal{L}_\mathrm{b}]=0$, so that we may simplify the above differential equation to
\begin{eqnarray}
\mathbb{P}\dot{\varrho}&\approx\mathbb{P}\bigl(\mathcal{L}_\mathrm{s}+\mathcal{L}_\mathrm{b}\bigr)\mathbb{P}\varrho+\int_0^\infty\rmd\tau\,\mathbb{P}\mathcal{L}_\mathrm{i}\mathbb{Q}e^{(\mathcal{L}_\mathrm{s}+\mathcal{L}_\mathrm{b})\tau}\mathcal{L}_\mathrm{i}e^{-\mathbb{P}(\mathcal{L}_\mathrm{s}+\mathcal{L}_\mathrm{b})\tau}\bigl(\rho_\mathrm{a}\otimes\Tr_\mathrm{a}\varrho\bigr)\\
&=\mathbb{P}\bigl(\mathcal{L}_\mathrm{s}+\mathcal{L}_\mathrm{b}\bigr)\mathbb{P}\varrho+\int_0^\infty\rmd\tau\,\mathbb{P}\mathcal{L}_\mathrm{i}\mathbb{Q}e^{(\mathcal{L}_\mathrm{s}+\mathcal{L}_\mathrm{b})\tau}\mathcal{L}_\mathrm{i}\bigl(\rho_\mathrm{a}\otimes e^{-\mathcal{L}_\mathrm{s}\tau}\Tr_\mathrm{a}\varrho\bigr).
\end{eqnarray}
We can now trace out mode `a,' thereby obtaining an approximate master equation for the reduced density matrix $\rho:=\Tr_\mathrm{a}\varrho$,
\begin{equation}
\dot{\rho}=\mathcal{L}_\mathrm{s}\rho_\mathrm{b}+\Tr_\mathrm{a}\int_0^\infty\rmd\tau\,\mathcal{L}_\mathrm{i}\mathbb{Q}e^{(\mathcal{L}_\mathrm{s}+\mathcal{L}_\mathrm{b})\tau}\mathcal{L}_\mathrm{i}\bigl(\rho_\mathrm{a}\otimes e^{-\mathcal{L}_\mathrm{s}\tau}\rho\bigr)\,.
\end{equation}
The integral can be evaluated by using the definition of $\mathcal{L}_\mathrm{i}$ given above. Upon applying the rotating-wave approximation, where quickly-rotating terms are neglected, and performing some algebra, the standard form
\begin{eqnarray}
\label{eq:EliminatedME}
\dot{\rho}&=\mathcal{L}_\mathrm{s}\rho-ig^2\left|\alpha\right|^2\im{S(\omega)+S(-\omega)}\bigl[\hat{b}^\dagger\hat{b},\rho\bigr]\nonumber\\
&\quad+g^2\left|\alpha\right|^2\bigl\{\re{S(\omega)}D\!\bigl[\hat{b}\bigr]\rho+\re{S(-\omega)}D\!\bigl[\hat{b}^\dagger\bigr]\rho\bigr\}
\end{eqnarray}
is recovered. In this expression, we have defined the spectral density of the cavity field to zeroth order in $g$, which can be expressed as
\begin{equation}
S(\omega):=\int_0^\infty\rmd\tau\,e^{i\omega\tau}\langle\hat{a}(t+\tau)\hat{a}^\dagger(t)\rangle=-\frac{1}{i(\Delta+\omega)-\kappa}.
\end{equation}
By comparing Eq.~\eref{eq:EliminatedME} to the standard master equation for a damped harmonic oscillator, we can identify an effective Hamiltonian
\begin{eqnarray}
\hat{H}_\mathrm{eff}&=\hat{H}_\mathrm{s}+g^2\left|\alpha\right|^2\im{S(\omega)+S(-\omega)}\hat{b}^\dagger\hat{b}\nonumber\\
&=\hat{H}_\mathrm{s}+\frac{2g^2\left|\alpha\right|^2\Delta\bigl(\Delta^2-\omega^2+\kappa^2\bigr)}{\bigl(\Delta^2-\omega^2-\kappa^2\bigr)^2+\bigl(2\Delta\kappa\bigr)^2}\hat{b}^\dagger\hat{b},
\end{eqnarray}
as well as the effective cooling and heating Liouvillians
\begin{equation}
\mathcal{L}_\mathrm{cool}:=\Biggl[\bigl(n+1\bigr)\gamma+\frac{g^2\left|\alpha\right|^2\kappa}{\bigl(\Delta+\omega\bigr)^2+\kappa^2}\Biggr]D\!\bigl[\hat{b}\bigr],
\end{equation}
and
\begin{equation}
\mathcal{L}_\mathrm{heat}:=\Biggl[n\gamma+\frac{g^2\left|\alpha\right|^2\kappa}{\bigl(\Delta-\omega\bigr)^2+\kappa^2}\Biggr]D\!\bigl[\hat{b}^\dagger\bigr].
\end{equation}

\section*{Appendix II: Equations of motion for the second-order moments}
The equations of motion for the second-order moments obtained from Eq.~\eref{eq:rhoeff2osc} are
\begin{eqnarray*}
\partial_t\expt{\hat{x}_1^2}=\omega\expt{\{\hat{x}_1,\hat{p}_1\}},\\
\partial_t\expt{\hat{x}_2^2}=\omega\expt{\{\hat{x}_2,\hat{p}_2\}},\\
\partial_t\expt{\hat{x}_1\hat{x}_2}=\partial_t\expt{\hat{x}_2\hat{x}_1}=\omega(\expt{\hat{x}_1\hat{p}_2}+\expt{\hat{x}_2\hat{p}_1}),\\
\partial_t\expt{\{\hat{x}_1,\hat{p}_1\}}=2\omega(\expt{\hat{p}_1^2}-\expt{\hat{x}_1^2})-2\Lambda\expt{\hat{x}_1\hat{x}_2}-(2\gamma+\bar{\gamma})\expt{\{\hat{x}_1,\hat{p}_1\}}\nonumber\\\hspace{3cm}-2\bar{\gamma}\expt{\hat{x}_1\hat{p}_2},\\
\partial_t\expt{\{\hat{x}_2,\hat{p}_2\}}=2\omega(\expt{\hat{p}_2^2}-\expt{\hat{x}_2^2})-2\Lambda\expt{\hat{x}_1\hat{x}_2}-(2\gamma+\bar{\gamma})\expt{\{\hat{x}_2,\hat{p}_2\}}\nonumber\\\hspace{3cm}-2\bar{\gamma}\expt{\hat{p}_1\hat{x}_2},\\
\partial_t\expt{\{\hat{x}_1,\hat{p}_2\}}=2\omega(\expt{\hat{p}_1\hat{p}_2}-\expt{\hat{x}_1\hat{x}_2})-2\Lambda\expt{\hat{x}_1^2}-2(2\gamma+\bar{\gamma})\expt{\hat{x}_1\hat{p}_2}\nonumber\\\hspace{3cm}-\bar{\gamma}\expt{\{\hat{x}_1,\hat{p}_1\}},\\
\partial_t\expt{\{\hat{x}_2,\hat{p}_1\}}=2\omega(\expt{\hat{p}_1\hat{p}_2}-\expt{\hat{x}_1\hat{x}_2})-2\Lambda\expt{\hat{x}_2^2}-2(2\gamma+\bar{\gamma})\expt{\hat{x}_2\hat{p}_1}\nonumber\\\hspace{3cm}-\bar{\gamma}\expt{\{\hat{x}_2,\hat{p}_2\}},\\
\partial_t\expt{\hat{p}_1^2}=-\omega\expt{\{\hat{x}_1,\hat{p}_1\}}-2\Lambda\expt{\hat{x}_2\hat{p}_1}-2(2\gamma+\bar{\gamma})\expt{\hat{p}_1^2}-2\bar{\gamma}\expt{\hat{p}_1\hat{p}_2}\nonumber\\\hspace{3cm}+\bigl[2\gamma(2n_1+1)+\bar{\gamma}(2\bar{n}+1)\bigr],\\
\partial_t\expt{\hat{p}_2^2}=-\omega\expt{\{\hat{x}_2,\hat{p}_2\}}-2\Lambda\expt{\hat{x}_1\hat{p}_2}-2(2\gamma+\bar{\gamma})\expt{\hat{p}_2^2}-2\bar{\gamma}\expt{\hat{p}_1\hat{p}_2}\nonumber\\\hspace{3cm}+\bigl[2\gamma(2n_2+1)+\bar{\gamma}(2\bar{n}+1)\bigr],\\
\partial_t\expt{\hat{p}_1\hat{p}_2}=\partial_t\expt{\hat{p}_2\hat{p}_1}=-\omega(\expt{\hat{x}_1\hat{p}_2}+\expt{\hat{p}_1\hat{x}_2})-\Lambda(\{\hat{x}_1\hat{p}_1\}+\{\hat{p}_2\hat{x}_2\})\nonumber\\
\hspace{3cm}-2(2\gamma+\bar{\gamma})\expt{\hat{p}_1\hat{p}_2}-\bar{\gamma}(\expt{\hat{p}_1^2}+\expt{\hat{p}_2^2})+\bar{\gamma}(2\bar{n}+1).
\end{eqnarray*}
Recall also that $\expt{[\hat{x}_j,\hat{p}_k]}=i\delta_{j,k}$, which gives $\partial_t\expt{[\hat{x}_j,\hat{p}_k]}=0$ $(j,k=1,2$).


\section*{References}
\begin{thebibliography}{99}
\bibitem{Dhar2008} A. Dhar, Adv. Phys. {\bf 57}, 457 (2008).
\bibitem{Plenio2004} M. B. Plenio, J. Hartley, and J. Eisert, New J. Phys. {\bf 6}, 36 (2004).
\bibitem{Galve2009} F. Galve and E. Lutz, Phys. Rev. A {\bf 79}, 032327 (2009).
\bibitem{Galve2010} F. Galve, G. L. Giorgi, and R. Zambrini, Phys. Rev. A {\bf 81}, 062117 (2010).
\bibitem{Asadian2013} A. Asadian, D. Manzano, M. Tiersch, and H. J. Briegel, Phys. Rev. E {\bf 87}, 012109 (2013).
\bibitem{Ghesquiere2013} A. Guesqui\`{e}res, I. Sinayskiy, and F. Petruccione, Phys. Lett. A {\bf 377}, 1682 (2013).
\bibitem{Manzano2013} G. Manzano, F. Galve, G. L. Giorgi, E. Hernandez-Garcia, and R. Zambrini, Sci. Rep. {\bf 3}, 1439 (2013).
\bibitem{Mari2013} A. Mari, A. Farace, N. Didier, V. Giovanetti, and R. Fazio, Phys. Rev. Lett. {\bf 111}, 103605 (2013).
\bibitem{Dhar2007} K. Saito, and A. Dhar, Phys. Rev. Lett. {\bf 99}, 180601 (2007).
\bibitem{Fogedby2014} H. C. Fogedby, A. Imparato, J. Stat. Mech. \textbf{2012}, P04005 (2012), H. C. Fogedby, A. Imparato, J. Stat. Mech. \textbf{2014}, P11011 (2014).
\bibitem{Kosloff2013} R. Kosloff, Entropy {\bf 15}, 2100-2128 (2013)
\bibitem{Strasberg2013} P. Strasberg, G. Schaller, T. Brandes, and M. Esposito, Phys. Rev. E {\bf 88}, 062107 (2013).
\bibitem{Esposito2010} M. Esposito, K. Lindenberg, and C. Van den Broeck, New J. Phys. {\bf 12}, 013013 (2010).
\bibitem{Eisert2004} J. Eisert, M. B. Plenio, S. Bose, and J. Hartley, Phys. Rev. Lett. {\bf 93}, 190402 (2004).
\bibitem{Heinrich2011} G. Heinrich, M. Ludwig, J. Qian, B. Kubala, and F. Marquardt, Phys. Rev. Lett. {\bf 107}, 043603 (2011).
\bibitem{Chang2011} D. E. Chang, A. H. Safavi-Naeini, M. Hafezi, and O. Painter, New J. Phys. {\bf 13}, 023003 (2011).
\bibitem{Tomadin2012} A. Tomadin, S. Diehl, M. D. Lukin, P. Rabl, and P. Zoller, Phys. Rev. A {\bf 86}, 033821 (2012).
\bibitem{Seok2012} H. Seok, L. F. Buchmann, S. Singh, and P. Meystre, Phys. Rev. A {\bf 86}, 063829 (2012).
\bibitem{Schmidt2012} M. Schmidt, M. Ludwig, and F. Marquardt, New J. Phys. {\bf 14}, 125005 (2012).
\bibitem{Xuereb2012} A. Xuereb, C. Genes, and A. Dantan, Phys. Rev. Lett. {\bf 109}, 223601 (2012).
\bibitem{Xuereb2013} A. Xuereb, C. Genes, and A. Dantan, Phys. Rev. A {\bf 88}, 053803 (2013).
\bibitem{Botter2013} T. Botter, D. W. C. Brooks, S. Schreppler, N. Brahms, and D. M. Stamper-Kurn, Phys. Rev. Lett. {\bf 110}, 153001 (2013).
\bibitem{Aspelmeyer2013} M. Aspelmeyer, T. J. Kippenberg, and F. Marquardt, arXiv:1303.0733 (2013).
\bibitem{Mari2012} A. Mari and J. Eisert, Phys. Rev. Lett. {\bf 108}, 120602 (2012).
\bibitem{Xuereb2014} A. Xuereb, C. Genes, G. Pupillo, M. Paternostro, and A. Dantan, Phys. Rev. Lett. {\bf 112}, 133604 (2014).
\bibitem{Zhang2014} K. Zhang, F. Bariani, and P. Meystre, Phys. Rev. Lett. {\bf 112}, 150602 (2014); Phys. Rev. A {\bf 90}, 023819 (2014).
\bibitem{Ian2014} H. Ian, arXiv:1402.3787 (2014).
\bibitem{Mari2014} A. Mari, A. Farace, and V. Giovanetti, arXiv:1407.8364 (2014).
\bibitem{Dechant2014} A. Dechant, N. Kiesel, and E. Lutz, arXiv:1408.4617 (2014).
\bibitem{Massel2012} F. Massel, S. U. Cho, J.-M. Pirkkalainen, P. J. Hakonen, T. T. Heikkil\"{a}, and M. A. Sillanp\"{a}\"{a}, Nat. Commun. {\bf 3}, 987 (2012).
\bibitem{Zhang2012} M. Zhang, G. S. Wiederhecker, S. Manipatruni, A. Barnard, P. McEuen, and M. Lipson, Phys. Rev. Lett. {\bf 109}, 233906 (2012).
\bibitem{Ludwig2013} M. Ludwig and F. Marquardt, Phys. Rev. Lett. {\bf 111}, 073603 (2013).
\bibitem{Bagheri2013} M. Bagheri, M. Poot, L. Fan, F. Marquardt, and H. X. Tang, Phys. Rev. Lett. {\bf 111}, 213902 (2013).
\bibitem{Kemiktarak2014} U. Kemiktarak, M. Durand, M. Metcalfe, and J. Lawall, Phys. Rev. Lett. {\bf 113}, 030802 (2014).
\bibitem{Schmidt2013} M. Schmidt, V. Peano, and F. Marquardt, arXiv:1311.7095 (2013).
\bibitem{Chesi2014} S. Chesi, Y.-D. Wang, and J. Twamley, arXiv:1402.0926 (2014).
\bibitem{Ciliberto13} 
S. Ciliberto, A. Imparato, A. Naert, M. Tanase,   Phys. Rev. Lett., {\bf 110}: 180601 (2013).
\bibitem{Pekola13}
J. V. Koski {\it et al.},  Nat. Phys. {\bf 9}, 644 (2013).
\bibitem{Seifert} U. Seifert, Rep. Prog. Phys. {\bf 75}, 126001 (2012).
\bibitem{Nicacio2014} F. Nicacio, A. Ferraro, A. Imparato, M. Paternostro, F. L. Semi\~{a}o, arXiv:1410.7604 (2014).
\bibitem{Wilsonrae2007} I. Wilson-Rae, N. Nooshi, W. Zwerger, and T. J. Kippenberg, Phys. Rev. Lett. {\bf 99}, 093901 (2007).
\bibitem{Marquardt2007} F. Marquardt, J. P. Chen, A. A. Clerk, and S. M. Girvin, Phys. Rev. Lett. {\bf 99}, 093902 (2007).
\bibitem{Genes2008} C. Genes, D. Vitali, P. Tombesi, S. Gigan, and M. Aspelmeyer, Phys. Rev. A {\bf 77}, 033804 (2008).
\bibitem{Dantan2008} A. Dantan, C. Genes, D. Vitali, and M. Pinard, Phys. Rev. A {\bf 77}, 011804 (2008).
\bibitem{Pinard2005} M. Pinard, A. Dantan, D. Vitali, O. Arcizet, T. Briant, and A. Heidmann, Europhys. Lett. {\bf 72}, 747 (2005).
\bibitem{Zhang2003} J. Zhang, K. Peng, and S. L. Braunstein, Phys. Rev. A {\bf 68}, 013808 (2003).
\bibitem{Gardiner2004} C. W. Gardiner and P. Zoller, {\it Quantum Noise}, 3rd ed. (Springer, 2004).
\bibitem{Thompson2008} J. D. Thompson, B. M. Zwickl, A. M. Jayich, F. Marquardt, S. M. Girvin, and J. G. E. Harris, Nature {\bf 452}, 72 (2008).
\bibitem{Arbadi2011} A. Arbadi, Y. M. Kang, C.-Y. Lu, E. Chow, and L. L. Goddard, Appl. Phys. Lett. {\bf 99}, 091105 (2011).
\bibitem{Eichenfield2009} M. Eichenfield, J. Chan, R. M. Camacho, K. J. Vahala, and O. Painter, Nature {\bf 462}, 78 (2009).
\bibitem{Brahms2012} N. Brahms, T. Botter, S. Schreppler, D. W. C. Brooks, and D. M. Stamper-Kurn, Phys. Rev. Lett. {\bf 108}, 133601 (2012).
\bibitem{Teufel2011} J. D. Teufel, T. Donner, Dale Li, J. W. Harlow, M. S. Allman, K. Cicak, A. J. Sirois, J. D. Whittaker, K. W. Lehnert, and R. W. Simmonds, Nature {\bf 475}, 359 (2011).
\bibitem{Chan2011} J. Chan, T. P. Mayer Alegre, A. H. Safavi-Naeini, J. T. Hill, A. Krause, S. Gr\"{o}blacher, M. Aspelmeyer, and O. Painter, Nature {\bf 478}, 89 (2011).
\bibitem{Purdy2014} T. P. Purdy, P.-L. Yu, N. S. Kampel, R. W. Peterson, K. Cicak, R. W. Simmonds, and C. A. Regal, arXiv:1406.7247 (2014).
\bibitem{Adiga2013} V. P. Adiga, R. De Alba, I. R. Storch, P. A. Yu, B. Ilic, R. A. Barton, S. Lee, J. Hone, P. L. McEuen, J. M. Parpia, and H. G. Craighead, Appl. Phys. Lett. {\bf 103}, 143103 (2013).
\bibitem{Bagci2013} T. Bagci, A. Simonsen, S. Schmid, L. G. Villanueva, E. Zeuthen, J. Appel, J. M. Taylor, A. S{\o}rensen, K. Usami, A. Schliesser, and E. S. Polzik, Nature {\bf 507}, 81 (2014).
\bibitem{Andrews2014} R. W. Andrews, R. W. Peterson, T. P. Purdy, K. Cicak, R. W. Simmonds, C. A. Regal, and K. W. Lehnert, Nature Phys. {\bf 10}, 421 (2013).
\bibitem{OConnell2010} A. D. O'Connell, M. Hofheinz, M. Ansmann, R. C. Bialczak, M. Lenander, E. Lucero, M. Neeley, D. Sank, H. Wang, M. Weides, J. Wenner, J. M. Martinis, and A. N. Cleland, Nature {\bf 464}, 697 (2010).
\bibitem{Pirkkalainen2013} J.-M. Pirkkalainen, S. U. Cho, J. Li, G. S. Paraoanu, P. J. Hakonen, and M. A. Sillanp\"{a}\"{a}, Nature {\bf 494}, 211 (2013).
\bibitem{Arcizet2006} O. Arcizet, P.-F. Cohadon, T. Briant, M. Pinard, and A. Heidmann, Nature {\bf 444}, 71 (2006).
\bibitem{Gigan2006} S. Gigan, H. R. B\"{o}hm, M. Paternostro, F. Blaser, G. Langer, J. B. Hertzberg, K. C. Schwab, D. B\"{a}uerle, M. Aspelmeyer, and A. Zeilinger, Nature {\bf 444}, 67 (2006).

\end{thebibliography}
\end{document}